\documentclass[prb,twocolumn,showpacs,preprintnumbers]{revtex4-1}
\usepackage{graphicx}
\usepackage{amsmath}
\usepackage{amssymb}
\usepackage{times}
\usepackage{color}
\usepackage{ulem}

\def\b{\beta}

\def\e{\epsilon}

\def\S{\Sigma}
\def\s{\sigma}
\def\t{\tau}
\def\w{\omega}
\def\ua{\uparrow}
\def\da{\downarrow}

\def\Vec#1{\mathbf #1}

\newcommand{\ksmall}{\tilde{\mathbf{k}}}

\begin{document}

\title{Cluster-size dependence in cellular dynamical mean-field theory}

\author{Shiro Sakai,$^{1,2}$ Giorgio Sangiovanni$^1$, Marcello Civelli$^3$, Yukitoshi Motome$^2$, Karsten Held$^1$, and Masatoshi Imada$^2$}

\affiliation{$^1$Institute for Solid State Physics, Vienna University of Technology, 1040 Vienna, Austria\\
$^2$Department of Applied Physics, University of Tokyo, Hongo,Tokyo 113-8656, Japan\\
$^3$Laboratoire de Physique des Solides, Universit\'e Paris-Sud, CNRS, UMR 8502, F-91405 Orsay Cedex, France
}
\date{\today}

\begin{abstract}
We examine the cluster-size dependence of the cellular dynamical mean-field theory (CDMFT) applied to the two-dimensional Hubbard model.
Employing the continuous-time quantum Monte Carlo method as the solver for the effective cluster model, we obtain CDMFT solutions for 4-, 8-, 12-, and 16-site clusters at a low temperature.
Comparing various periodization schemes, which are used to construct the infinite-lattice quantities from the cluster results, we find that the cumulant periodization yields the fastest convergence for the hole-doped Mott insulator where the most severe size dependence is expected.
We also find that the convergence is much faster around $(0,0)$ and $(\frac{\pi}{2},\frac{\pi}{2})$ than around $(\pi,0)$ and $(\pi,\pi)$.
The cumulant-periodized self-energy seems to be close to its thermodynamic limit already for a 16-site cluster in the range of parameters studied.
The 4-site results remarkably agree well with the 16-site results, indicating that the previous studies based on the 4-site cluster capture the essence of the physics of doped Mott insulators.
\end{abstract}
\maketitle

\section{INTRODUCTION}

A range of anomalies observed in the normal state of high-$T_\text{C}$ cuprates indicates a momentum-space differentiation of the electronic structure.
For instance, the pseudogap \cite{ts99} and Fermi arc \cite{nd98,dh03} observed by the angle-resolved photoemission spectroscopy suggest a gap in the single-particle excitation spectra around antinodal points [i.e., ($\pi,0$) and its symmetrically related points in the Brillouin zone] while there is a metallic spectrum around nodal points [i.e., ($\frac{\pi}{2},\frac{\pi}{2}$) and its symmetrically related points].
Since undoped cuprates are considered Mott insulators \cite{a87}, these findings have directed the attention to doped Mott insulators where strong electronic correlations may cause unprecedented metallic states.
The simplest play-ground model which describes the doped Mott insulators is the Hubbard model in two dimensions.
In fact, various numerical calculations for the model suggest a momentum-space differentiation of electronic properties at small doping,\cite{st04,cc05,kk06,sk06,fc09,wg09,c09,sm09,sm10} similarly to the experimental results on cuprates.
Nevertheless, many fundamental questions remain unresolved on how the metal-Mott insulator transition takes place in two dimensions:
How does the Fermi surface evolve while approaching the Mott insulator?
Is there a Fermi arc or Fermi pocket in the underdoped region? 
Is the length of this Fermi arc/pocket going to zero while approaching the Mott insulator? 
Or/and is the weight of the quasiparticle excitations fading away at the Mott transition?
Is there a quantum critical point and/or non-Fermi liquid phase in between the Fermi liquid and the Mott insulator?
Is the Luttinger sum rule fulfilled in the whole doping range?

\begin{figure}[t]
\center{
\includegraphics[width=0.50\textwidth]{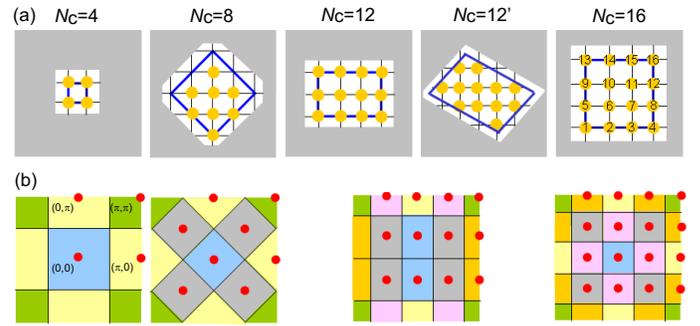}}
\caption{(Color online). 
         (a) Cluster geometries employed in the present CDMFT study.
         For comparison, (b) shows the typical momentum patches 
         used in DCA for the same cluster sizes, where the dots
         denote the central momenta in each patch.}
\label{fig:cluster}
\end{figure}

To address these issues, we need a method that can describe sufficiently well the momentum dependence of the electronic structure. We also need to treat on the same footing the low and high energy scales, since the electronic structure is reconstructed on a wide energy range in Mott-related phenomena. 
Schemes which have been developed to fulfill these requirements are for instance the cluster extensions of the dynamical mean-field theory (DMFT) \cite{gk96}, such as the cellular DMFT (CDMFT) \cite{ks01} and the dynamical cluster approximation (DCA).\cite{mj05}
Both theories map the Hubbard model onto an effective model consisting of an interacting small cluster and non-interacting infinite bath. CDMFT defines the effective model in real space while the DCA defines the effective model in momentum space (see Fig.~\ref{fig:cluster} for an illustration).
In both theories, the resolution in momentum space is limited by the cluster size, so that a high momentum resolution requires a large cluster, intractable with present computational resources.

The advantage of DCA is that it keeps the translational symmetry of the original lattice. This allows one to define cluster momenta by partitioning 
the Brillouin zone into several patches [Fig.~\ref{fig:cluster}(b)]. The momentum dependence of the self-energy within each patch is neglected.
Then, the coarse-grained Green's function, which is employed in the
self-consistency loop, is defined by averaging over all momenta within
each patch (e.g., including momentum on the Fermi surface and far away
from it).
Hence, in the parameter region where the momentum-space differentiation is crucial, the analysis with a small cluster may lose important information.
For example, a recent systematic study on the cluster-size dependence in the DCA\cite{gf10} has shown that the nodal-antinodal momentum-space differentiation is not clearly seen with the 4-site cluster patched as in the left panel in Fig.~\ref{fig:cluster}(b), while it is seen with another choice of the momentum patch (and with larger clusters).

On the other hand, CDMFT is performed by defining a real-space finite-size cluster. It explicitly takes into account short-range correlations within the cluster.
This scheme does not need the average over momentum, but instead, it
breaks the translational invariance, which has to be restored only at the end by deriving lattice quantities from the cluster ones through a periodization procedure. This latter problem has been analyzed by some of the authors \cite{c09,sm09,sm10} using the 2$\times$2-cluster CDMFT. The results suggest that a normal Fermi liquid, realized at high doping, evolves by reducing doping toward a Mott insulator through (at least) two quantum phase transitions to non-Fermi liquid phases by topological changes of the Fermi surface and the appearance of a surface of zeros of Green's function (i.e., poles of the self-energy).
The non-Fermi liquid phases have hole-pocket Fermi surfaces at low doping and show arclike spectra at finite temperatures. 

In order to check how robust early CDMFT results on the 2$\times$2 and 2$\times$1 clusters \cite{st04,cc05,kk06,sk06,c09,sm09,sm10,ks08} are against the increase of cluster size $N_\text{C}$, larger cluster studies have been highly desired.
A systematic study by increasing cluster size helps also to identify the best periodization scheme which could reach the thermodynamic limit in the fastest way.\cite{bk02}
This would allow us to understand to what extent small cluster calculations (still accessible by present computational limits) capture the right physical results, once implemented with the most suitable periodization scheme.

In this paper, we extend CDMFT up to 16-site cluster, employing the continuous-time quantum Monte Carlo method (CTQMC) \cite{gm11} as the solver for the effective models.
By comparing the solutions for the 4-, 8-, 12-, and 16-site clusters illustrated in Fig.~\ref{fig:cluster}(a), we systematically study the cluster-size dependence of the single-particle Green's function and the self-energy obtained through various periodization schemes.
We find that the cumulant periodization \cite{sk06} gives the fastest convergence against the cluster size for slightly doped Mott insulators, where the size dependence is severest due to the strong momentum dependence of the self-energy.
We also find that the convergence is strongly momentum-dependent: It is faster around $(0,0)$ and $(\frac{\pi}{2},\frac{\pi}{2})$, where the results for 16 sites seem nearly converged, than around $(\pi,0)$ and $(\pi,\pi)$.
Remarkably the 2$\times$2 cluster results are found to agree well with the 4$\times$3 and 4$\times$4 cluster results.
In particular, the location of the Fermi arc and the zero surface of Green's function at the Fermi level does not depend significantly on the cluster size.
This suggests that previous CDMFT results obtained by using a 4-site cluster \cite{cc05,sk06,kk06,c09,sm09,sm10} capture an essential physics of the slightly doped Mott insulators.

The paper is organized as follows.
In Sec.~\ref{sec:method} we briefly introduce the Hubbard model, CDMFT, and various periodization schemes
(in Appendix A the different periodization schemes are discussed from a different perspective).
The numerical results obtained by CDMFT are discussed in Sec.~\ref{sec:result}:
We first study the inhomogeneity and locality of various cluster quantities, i.e., self-energy, cumulant, and Green's function, and then discuss the periodized quantities for various square clusters whose edges are parallel to the lattice vectors, with a focus on the low-energy electronic structure. The results for tilted clusters of $N_\text{C}=8$ and 12 are presented in Appendix B, where the effect of the cluster geometry is discussed.
We summarize the results in Sec.~\ref{sec:summary}.

\section{MODEL AND METHOD} \label{sec:method}

We study the two-dimensional single-band Hubbard model on a square lattice.
The Hamiltonian reads
\begin{align}
H= \sum_{\Vec{k}\s}\e(\Vec{k})c_{\Vec{k}\s}^\dagger c_{\Vec{k}\s}^{\phantom {\dagger}}
-\mu \sum_{i\s}n_{i\s}+ U\sum_{i}n_{i\ua}n_{i\da},
\label{eq:hubbard}
\end{align}
where $c_{\Vec{k}\s}^{\phantom {\dagger}}$ $(c_{\Vec{k}\s}^\dagger)$ annihilates (creates) an electron with spin $\s$ and momentum $\Vec{k} = (k_x, k_y)$, 
$c_{i\s}^{\phantom {\dagger}}$ $(c_{i\s}^\dagger)$ is its Fourier component at site $i$, and
$n_{i\s}\equiv c_{i\s}^\dagger c_{i\s}^{\phantom {\dagger}}$.
Here, $U$ represents the onsite Coulomb repulsion, $\mu$ the chemical potential, and 
\begin{align}
\e(\Vec{k})\equiv -2t(\cos k_x + \cos k_y) -4t' \cos k_x\cos k_y,
\label{eq:disp}
\end{align}
where $t$ $(t')$ is the (next-)nearest-neighbor transfer integral. 
We solve the model within CDMFT,\cite{ks01} restricting the solution to the paramagnetic state to focus on Mott physics.
We adopt a set of parameters appropriate for hole-doped cuprates: $t'=-0.2t$ and $U=8t$,\cite{hs90} for which the CMDFT solution is a Mott insulator at half filling ($n=1$) for all cluster sizes we consider. We study three values of hole doping, 1, 3, and 5\% and fix the temperature to $T=0.06t$.
Because of the well-known fermionic sign problem, which becomes severer with decreasing filling and temperature, we cannot reach larger dopings and/or lower temperatures for our largest cluster $N_\text{C}=16$.
In CDMFT the infinite lattice quantum problem (\ref{eq:hubbard}) is
mapped onto an effective cluster model,
\begin{align}
S_{\text{eff}}= 
-\int_{0}^{\beta} \, d\tau \, d\tau' \sum_{i,j}^{N_\text{C}}\sum_{\s} \,  
c_{i\s}^\dagger(\tau)\, \mathcal{G}_{0,ij}^{-1}(\tau-\tau')\, c_{j\s}^{\phantom {\dagger}}(\tau') \nonumber \\
+ \int_{0}^{\beta} \, d\tau \, U \sum_{i}^{N_\text{C}} \,  n_{i\ua}(\tau) n_{i\da}(\tau),
\label{eq:action}
\end{align}
consisting of an $N_\text{C}$-site cluster embedded in a bath of non-interacting fermions, which is described by a dynamical Weiss field matrix $\mathcal{\hat{G}}_{0}$ at the inverse temperature $\b=\frac{1}{T}$.

We solve the cluster model by means of CTQMC.\cite{gm11}
We choose the interaction-expansion variant \cite{rh99,al06,gw08} which is the most suited approach to address large clusters.
Recent progress \cite{nm09} in the updating algorithm in the CTQMC enables us to reach much lower temperatures than those in previous studies.
The CDMFT-CTQMC approach incorporates all the correlations within the cluster, so that it converges to the exact solution as $N_\text{C}\rightarrow \infty$.
Here we study 4-, 8-, 12-, and 16-site clusters with the geometries illustrated in Fig.~\ref{fig:cluster}(a).
The increasing size of the Green's function matrix as well as the increasing negative signs of the QMC samples prevents us from studying even larger clusters.

The effective cluster model is subject to the CDMFT self-consistency condition which relates the cluster one-particle Green's functions 
$G_{ij}^\text{C}(i\omega_n)=\,-\int d\tau e^{i\omega_n\tau }\,\langle T_{\tau }c_{i}(\tau) c_{j}^{\dagger}(0)\rangle $ 
to the lattice Green's function of the original model,
\begin{eqnarray}
\hat{G}^\text{C}(i\omega_n) &=& \frac{N_{\text{C}}}{(2\pi)^2} \int_\text{RBZ}
 \hat{G}(\widetilde{\mathbf{k}}, i \omega_n)d\widetilde{\Vec{k}}, \label{eq:Gc} \\
\hat{G}(\widetilde{\mathbf{k}}, i \omega_n) &=&\left[ \left( i\omega_n+\mu \right) \hat{I}-\hat{t} (\widetilde{\mathbf{k}})- \hat{\Sigma}%
(i\omega_n)\right] ^{-1}, \label{eq:selfcon}
\end{eqnarray}
where $\hat{\Sigma}(i\omega_n)=\mathcal{\hat{G}}_0 (i\omega _{n})^{-1}-\hat{G}^\text{C}(i\omega_n)^{-1}$ (all being matrices with respect to cluster sites).
Here, ${t}_{ij}(\widetilde{\mathbf{k}})=\sum_{\mathbf{K}}
e^{-i\left( \mathbf{K}+ \widetilde{%
\mathbf{k}}\right) \cdot \left( \mathbf{r}_{i}-\mathbf{r}_{j}\right)
}\epsilon ({\mathbf{K+}\widetilde{\mathbf{k}}})$ 
is the single-electron part of the Hamiltonian written in the reduced Brillouin zone (RBZ) of the cluster, $\widetilde{\mathbf{k}}$ is the wave vector in the RBZ, $\mathbf{K}$ the reciprocal vector of the cluster,
and $\mathbf{r}_i$ the intra-cluster vector coordinate.

So far CDMFT has been in most cases employed for 2$\times$2 (or smaller) cluster calculations. Some exceptions are a systematic analysis with Lanczos methods for one-dimensional clusters up to eight sites,\cite{ks08} a QMC analysis of the cluster-size dependence in one and two dimensions,\cite{kk06} a study about the different geometries in $d=2$ in Ref.~\onlinecite{ic09}, and 8- and 16-site calculations performed by some of the authors.\cite{sm10} 
As we already mentioned, CDMFT clusters break the translational
 symmetry of the lattice; they are defined with open boundary
 conditions.
This means that, unlike in DCA, cluster momenta are no longer good quantum numbers. 
This is not a problem as long as one stays in the basis defined by the good quantum numbers of CDMFT, namely, the irreducible representations of the point group of the cluster.\cite{ks08}  
However, if one wants to compare with, say, angle-resolved photoemission experiments, $\Vec{k}$-resolved spectral functions are needed.
Several ``estimators'' for lattice quantities have been proposed and in the recent literature there has been an intensive discussion on what is the best strategy to produce good $\Vec{k}$-resolved spectral functions.\cite{kk06,sk06,c09,sm10,bk02}
There are two main issues: one would like the cluster quantity that has to be periodized (i) to be as localized as possible, in order to minimize the impact of the approximation of neglecting inter-cluster correlations and (ii) to be as homogeneous as possible, so that translational invariance is (almost) fulfilled.

If criteria (i) and (ii) are both met for the cluster quantity $\hat{Q}^{\text{C}}(i\w_n)$, it is useful to define the corresponding lattice quantity by the Fourier expansion, 
\begin{align} 
 Q^{\text{L}}(\Vec{k},i\w_n)=\frac{1}{N_{\text{C}}}\sum_{i,j=1}^{N_\text{C}}
                      Q_{ij}^{\text{C}}(i\w_n)
                       \operatorname{e}^{i\Vec{k}\cdot(\Vec{r}_i-\Vec{r}_j)},
\label{eq:periodize}
\end{align}
truncated by the cluster size, since the longer-range terms would be negligible.
Here $\Vec{k}$ is defined on the entire Brillouin zone of the original lattice,  and, as introduced above, $\Vec{r}_{i}$ and $\Vec{r}_{j}$ are the real-space vectors specifying the cluster sites $i$ and $j$, respectively.\cite{footnote}

Two different choices for $Q$ have been mainly proposed so far: The cluster self-energy $\hat{\S}$ (Ref.~\onlinecite{cc05}) and the cumulant $\hat{M}\equiv[i\w_n+\mu-\hat{\S}]^{-1}$ (Ref.~\onlinecite{sk06}).
In the following, we call the two schemes $\Sigma$ periodization and $M$ periodization, respectively.
It is expected that in the weak-coupling regime the self-energy is small and the cumulant is well localized so that both choices can be appropriate.
On the other hand, in the strong-coupling regime, while the self-energy is large and nonlocal, the cumulant is well localized so that $Q=M$ is expected to be a more appropriate choice.\cite{sk06,sm10}
For $Q=\S$, according to  Eq.~(\ref{eq:periodize}), the lattice Green's function is given by
\begin{align}
G^\text{L}(\Vec{k},i\w_n)=[i\w_n+\mu-\e(\Vec{k})-\S^{\text{L}}(\Vec{k},i\w_n)]^{-1},\label{eq:G_S}
\end{align}
while, for $Q=M$, the lattice Green's function is given by 
\begin{align}
G^\text{L}(\Vec{k},i\w_n)=[M^{\text{L}}(\Vec{k},i\w_n)^{-1}-\e({\Vec{k}})]^{-1}, \label{eq:G_M}
\end{align}
and the self-energy reads
\begin{align}
\S^\text{L}(\Vec{k},i\w_n)=i\w_n+\mu-M^{\text{L}}(\Vec{k},i\w_n)^{-1}.
\end{align}

We can also use Green's function in order to build the translational-invariant object. We call this procedure $G$ periodization.
In this case, however, we have to modify Eq.~(\ref{eq:periodize}).
Indeed Eq.~(\ref{eq:periodize}) with $\hat{Q}^\text{C}=\hat{G}^\text{C}$ does not reproduce the correct lattice dispersion, as shown in Appendix A.
A prescription for this was previously proposed in Ref.~\onlinecite{sp00} and it has been already used in CDMFT studies, in particular in Ref.~\onlinecite{kk06}.
The $G$ periodization formula reads
\begin{align}
 G^{\text{L}}(\Vec{k},i\w_n)=\frac{1}{N_{\text{C}}}\sum_{i,j=1}^{N_\text{C}}
          & G_{ij}(\widetilde{\mathbf{k}},i\w_n) \, \operatorname{e}^{i\Vec{k}\cdot(\Vec{r}_i-\Vec{r}_j)},
\label{eq:Gscheme}
\end{align}
where $G_{ij}(\widetilde{\mathbf{k}}, i \omega_n)$ with $\widetilde{\mathbf{k}}=\mathbf{k}\ \text{modulo}\ \mathbf{K}$ is defined in Eq.~(\ref{eq:selfcon}).
In this case we obtain the lattice self-energy through
\begin{align}
\S^\text{L}(\Vec{k},i\w_n)=i\w_n+\mu-\e(\Vec{k})-G^{\text{L}}(\Vec{k},i\w_n)^{-1}.
\end{align}

In a large-frequency expansion (see Appendix A), these three periodization schemes give the same lattice Green's function up to the second order.
We expect, however, that the $M$ and $G$ schemes are closer to each other than to the $\S$ scheme near the Mott insulator (see again Appendix A for details).

A difference between the three periodization schemes becomes prominent only at low energy.
Hence, in the next section, we focus on low-energy behaviors.
We first compare the locality and inhomogeneity of the cluster quantities $Q^\text{C}$, which determine the accuracy of the periodization (\ref{eq:periodize}).
Even though Eq.~(\ref{eq:Gscheme}) in the $G$ periodization is not a truncated Fourier transformation, the cluster Green's function (\ref{eq:Gc}) is 
still relevant because it gives the average of Green's function (\ref{eq:selfcon}) over RBZ.
Therefore, we examine the locality and homogeneity of $M^\text{C}$, $\S^\text{C}$, and $G^\text{C}$ in the next section.

We will show that, in the parameter region of lightly hole-doped cuprates, the cumulant is at the same time well localized and fairly homogeneous over the cluster so that it certainly represents the best choice as an ``estimator'' of lattice quantities. Compared to DCA the advantage is that $\Vec{k}$-resolved quantities are not smeared out because the averages within each momentum patch of the Brillouin zone employed in DCA are not taken in CDMFT.
The disadvantage is that the periodization is not unique and it involves an artificial average of cluster quantities with the same lattice vector.

In the next section we will present results for different cluster sizes and geometries. The 2$\times$2 cluster is a bit special since, by symmetry, all sites are equivalent so that the quantities are already homogeneous over the cluster and the concept of ``cluster momenta'' is still meaningful. Another way to see this is that, since the cluster is entirely made of ``surface sites'', it fulfills periodic boundary conditions. 
For convenience's sake, in the following sections, we abbreviate $M^\text{L}$, $\S^\text{L}$, and $G^\text{L}$ with $M$, $\S$, and $G$, respectively.



\section{RESULT AND DISCUSSION}\label{sec:result}
\subsection{Inhomogeneity of cluster quantities} \label{ssec:ih}

\begin{figure}[t]
\center{
\includegraphics[width=0.48\textwidth]{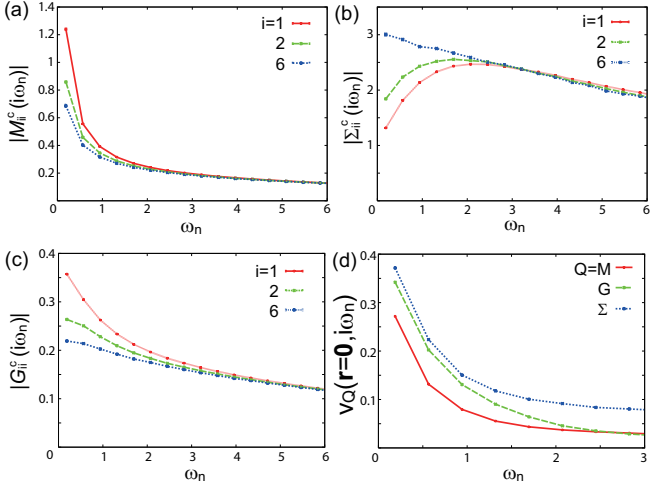}}
\caption{(Color online). 
        Absolute values of local components of 
        (a) the cluster cumulant ($Q=M$), (b) self-energy ($Q=\Sigma$)  and 
        (c) Green's function ($Q=G$) against the Matsubara frequency 
        for $N_\text{C}=16$ and $n=0.95$.
        The index $i$ denotes the cluster sites numbered in 
        the right-most panel of Fig.~\ref{fig:cluster}(a).
        The errorbars are below the symbol size, except for those at the 
        lowest Matsubara frequency.
        (d) Comparison of the normalized deviations defined 
        in Eq.~(\ref{eq:var}), which shows that the cumulant expansion is 
        the most homogeneous.}
\label{fig:ih_local}
\end{figure}

\begin{figure}[t]
\center{
\includegraphics[width=0.48\textwidth]{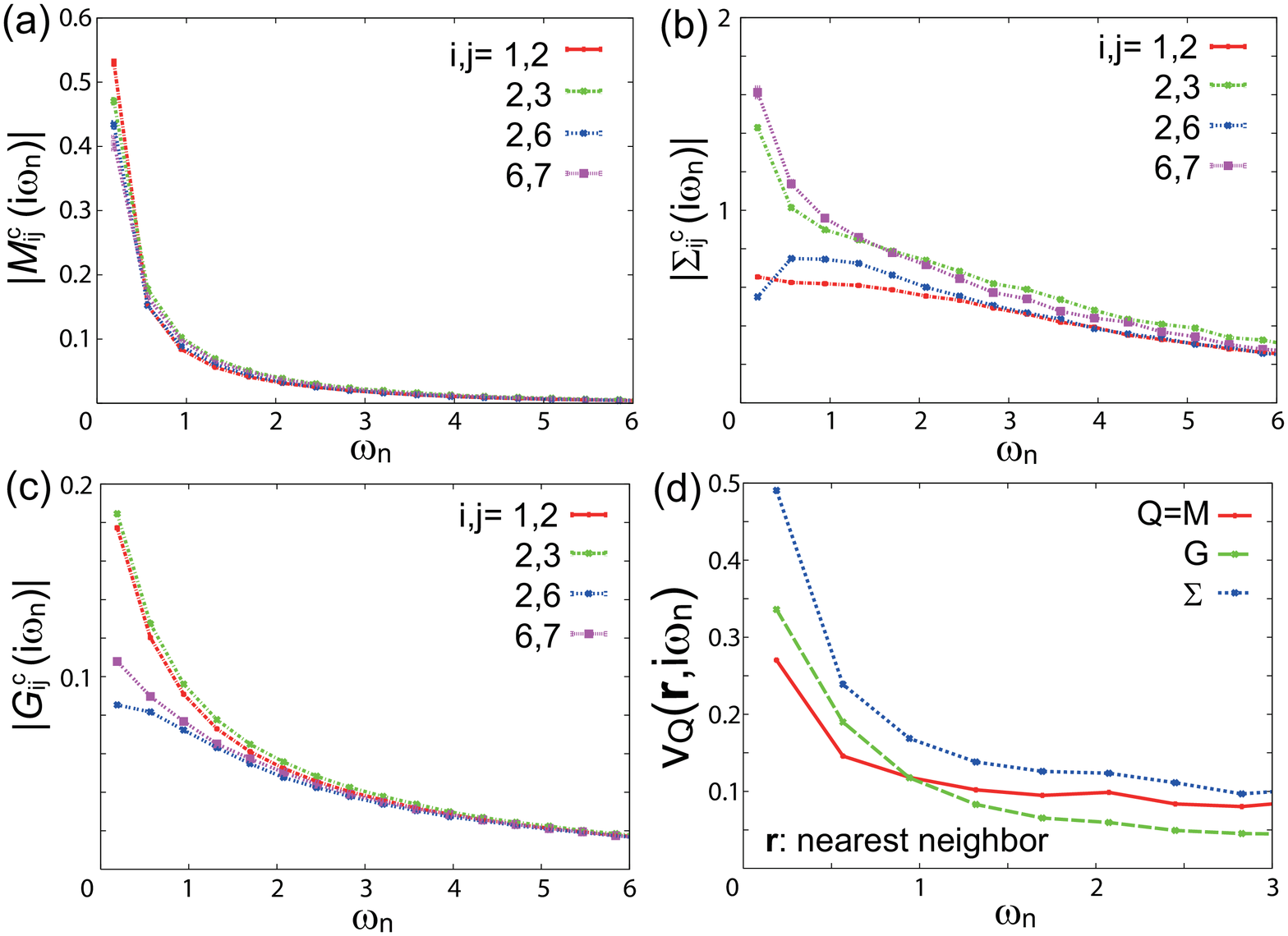}}
\caption{(Color online). 
        Same as Fig.~\ref{fig:ih_local} but for the nearest-neighbor components.}
\label{fig:ih_nn}
\end{figure}

\begin{figure}[t]
\center{
\includegraphics[width=0.48\textwidth]{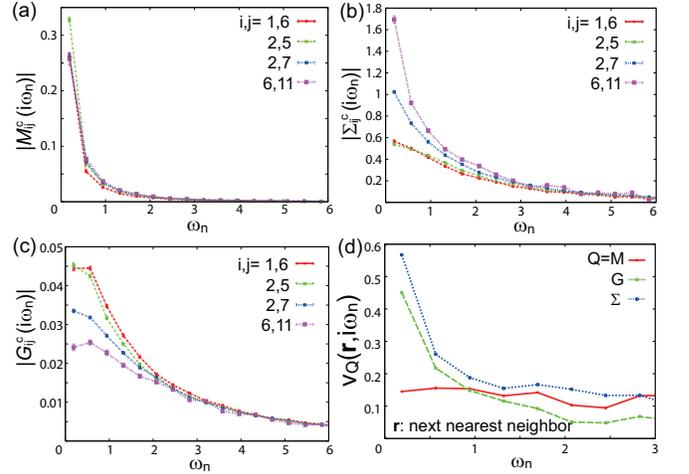}
}
\caption{(Color online). 
        Same as Fig.~\ref{fig:ih_local} but for the next-nearest-neighbor components.}
\label{fig:ih_nnn}
\end{figure}

We start with the inhomogeneity in the cluster quantities.
As we noted in the previous section, the inhomogeneity is present because the CDMFT violates the translational symmetry of the original lattice.
In the 4$\times$4 cluster, for example, the cluster sites, 
numbered in the right-most panel in Fig.~\ref{fig:cluster}(a),
 are categorized into three different symmetry groups: $\{1,4,13,16\}$, $\{6,7,10,11\}$, and $\{2,3,5,8,9,12,14,15\}$.
We pick up one site from each group and plot in Figs.~\ref{fig:ih_local}(a)-(c) the absolute value of the local cluster quantities, $M_{ii}^\text{C}$, $\S_{ii}^\text{C}$, and $G_{ii}^\text{C}$, against the Matsubara frequency for a hole-doped ($n=0.95$) Mott insulator.
The cluster quantities indeed depend on the group:
While the dependence is small at high energy, it is more pronounced at low energy.

An important finding in Fig.~\ref{fig:ih_local} is that the inhomogeneity occurs differently among $M^\text{C}$, $\S^\text{C}$, and $G^\text{C}$.
To quantify it, we define a normalized deviation of the cluster quantities at each real-space vector $\Vec{r}$ by 
\begin{align}
 v_{Q}(\Vec{r})\equiv \frac{1}{|\bar{Q}^\text{C}(\Vec{r})|}
               \sqrt{\frac{1}{N_\Vec{r}}
                    \sum_{i,j\in \text{C};\Vec{r}_i-\Vec{r}_j=\Vec{r}}
                    [Q_{ij}^\text{C}-\bar{Q}^\text{C}(\Vec{r})]^2},
              \label{eq:var}
\end{align}
where the average $\bar{Q}^\text{C}$ is defined by 
\begin{align}
 \bar{Q}^\text{C}(\Vec{r})\equiv 
 \frac{1}{N_\Vec{r}}
 \sum_{i,j\in \text{C};\Vec{r}_i-\Vec{r}_j=\Vec{r}}Q_{ij}^\text{C},
 \label{eq:av}
\end{align}
and $N_\Vec{r}$ is the number of pairs $(i,j)$ satisfying the conditions $i,j\in \text{C}$ and $\Vec{r}_i-\Vec{r}_j=\Vec{r}$.
Plotting the local ($\Vec{r}=\Vec{0}$) component for each of $M^\text{C}$, $\S^\text{C}$, and $G^\text{C}$ in Fig.~\ref{fig:ih_local}(d), we find that the normalized deviation is the smallest in $M^\text{C}$, in particular, in the wide range of the relevant low-frequency region. This means that the cumulant is most homogeneous within the cluster.

Figures \ref{fig:ih_nn}(a)-(c) and \ref{fig:ih_nnn}(a)-(c) plot the nearest-neighbor and next-nearest neighbor components, respectively.
These are the main sources of momentum dependence in the periodized quantities.
We again see the inhomogeneity occurring in different ways in $M^\text{C}$, $\S^\text{C}$, and $G^\text{C}$, and that the corresponding normalized deviations, shown in Figs.~\ref{fig:ih_nn}(d) and \ref{fig:ih_nnn}(d), are the smallest in $M^\text{C}$ at low energy.
Note that a relatively large fluctuation in the normalized deviations at high energy is due to the small absolute values of the cluster quantities.

The above results indicate that the inhomogeneity plays the weakest role for the $M$ periodization scheme in the hole-doped Mott insulator.

\subsection{Locality of cluster quantities} \label{ssec:locality}

\begin{figure}[t]
\center{
\includegraphics[width=0.48\textwidth]{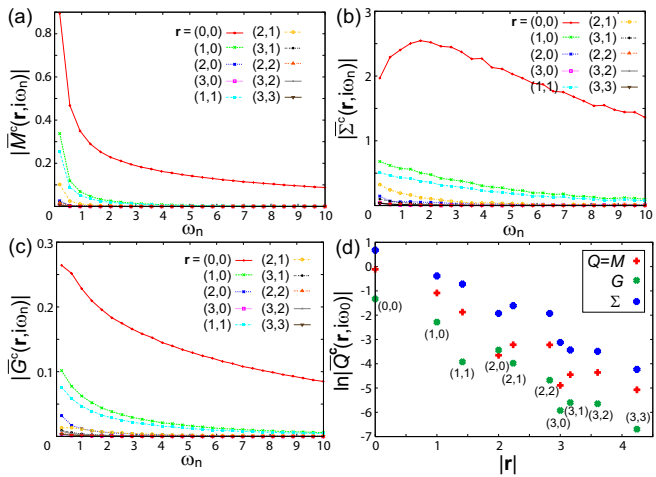}}
\caption{(Color online). Averaged cluster quantities $\bar{Q}^\text{C}$ for 
         $N_\text{C}=16$ and $n=0.95$ plotted against the Matsubara frequency.
        (a) The cluster cumulant, (b) self-energy, and (c) Green's function.
        (d) $\ln|\bar{Q}^\text{C}|$ at the lowest Matsubara frequency $\w_0$ 
        plotted against the Euclidean distance $|\Vec{r}|$.
        }
\label{fig:locality}
\end{figure}

We next discuss how the cluster quantities decay with the real-space distance.
Figures \ref{fig:locality}(a)-(c) plot the average quantities, $\bar{M}^\text{C}$, $\bar{\S}^\text{C}$, and $\bar{G}^\text{C}$, defined in Eq.~(\ref{eq:av}) at various $\Vec{r}$ against the Matsubara frequency.
We see that $(0,0)$, $(1,0)$, and $(1,1)$ components, which are within the 2$\times$2 cluster, are much larger than the other components for all kinds of cluster quantities.
 In particular in $\bar{M}^\text{C}$ and $\bar{G}^\text{C}$, even the $(1,1)$ component, the smallest one in the 2$\times$2 cluster, is always more than two times larger than the longer-range components. 
This implies that the longer-range terms play a less important role in the periodization procedure, above all within the $M$ scheme. 

In general, Green's function at a finite temperature should decay exponentially with $|\Vec{r}|$ at long distances, and cross over to a power-law decay at short distances in a metallic state. 
To explore the decay, we plot in Fig.~\ref{fig:locality}(d) $\ln|\bar{Q}^\text{C}|$ at the lowest Matsubara frequency $\w_0$ against the Euclidean distance $|\Vec{r}|$.
Although the data points fluctuate depending on direction, the decay seems to approximately follow an exponential decay at large distances and to cross over to a more moderate slope at short distances.
This behavior suggests that we are looking at the thermodynamic behavior already by the 4$\times$4 cluster at the present temperature.

When the temperature is reduced below 0.06$t$, the power-law decay expected to continue at longer distances in metals is, of course, not properly captured by the 4$\times$4 cluster.  Since the energy resolution is set by the lowest Matsubara frequency $\pi T \sim 0.2t$, the result suggests that the 4$\times$4 cluster well describes the thermodynamic behavior within this energy resolution (or at $T \ge 0.06t$). For instance, since the real and imaginary parts of Green's function have the denominator $\sim 1/[(\pi T)^2+ (v_\text{F}k)^2$], the correlation length of the exponential decay in distance is $\xi \sim v_\text{F}/\pi T$, where $v_\text{F}$ is the renormalized Fermi velocity.  In the underdoped region, the renormalization factor is suppressed and $v_\text{F}$ may easily be of the order of 0.1, which is consistent with the behavior in Fig.~\ref{fig:locality}(d).  Because of the proximity to the Mott insulator, the correlation length (or coherence length) of Green's function is suppressed comparable to the lattice constant even at fairly low temperatures like 0.06$t$.  Nevertheless, the characteristics such as the pseudogap, momentum differentiations and the Fermi arc formation are well developed in this temperature range.

\subsection{Momentum dependence of periodized quantities} \label{ssec:kdep}

\begin{figure}[t]
\center{
\includegraphics[width=0.48\textwidth]{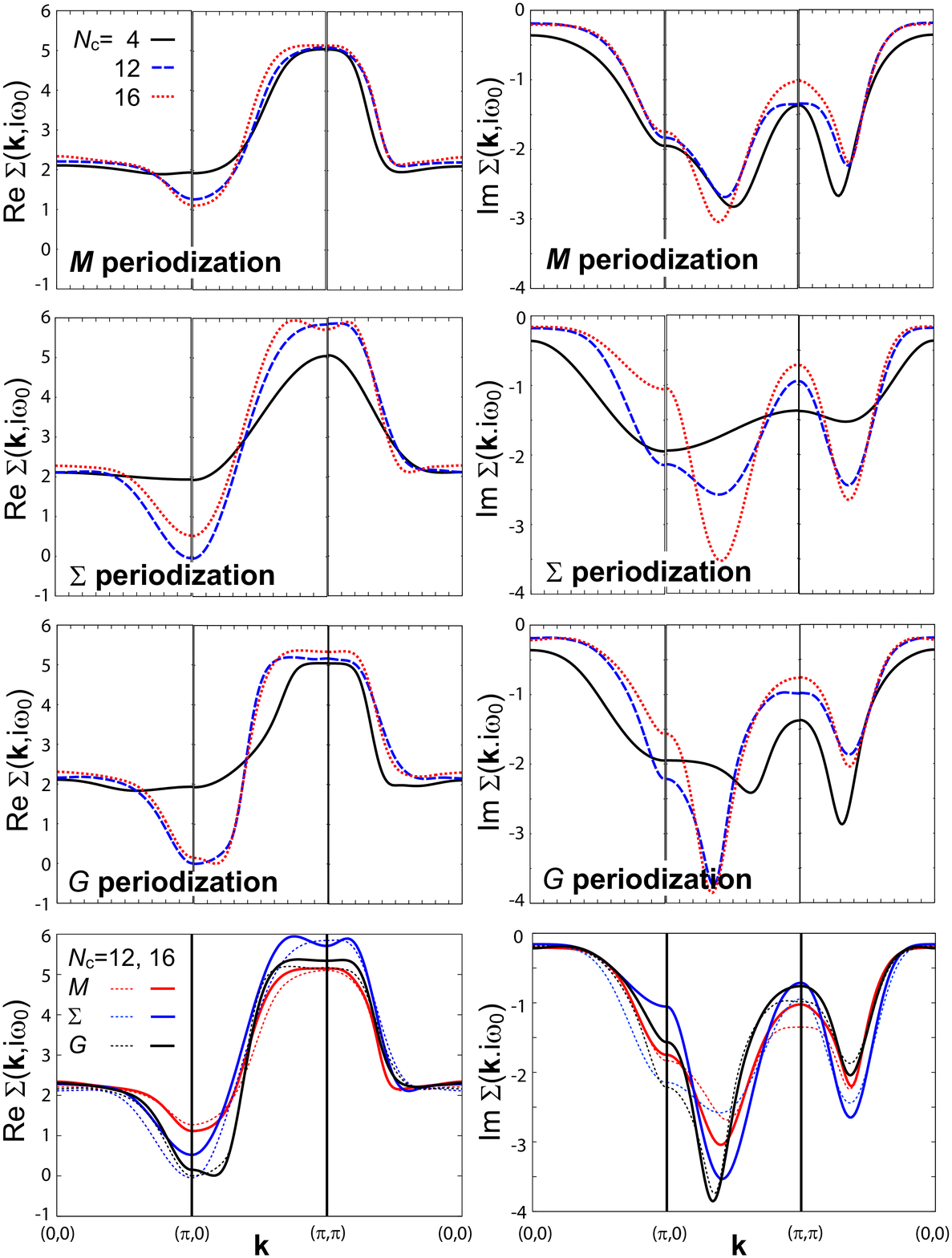}}
\caption{(Color online). Self-energies at the lowest Matsubara frequency, 
         obtained by the $M$, $\S$, and $G$ periodizations for various 
         square clusters at $n=0.95$.
        }
\label{fig:kdep}
\end{figure}

\begin{figure}[t]
\center{
\includegraphics[width=0.48\textwidth]{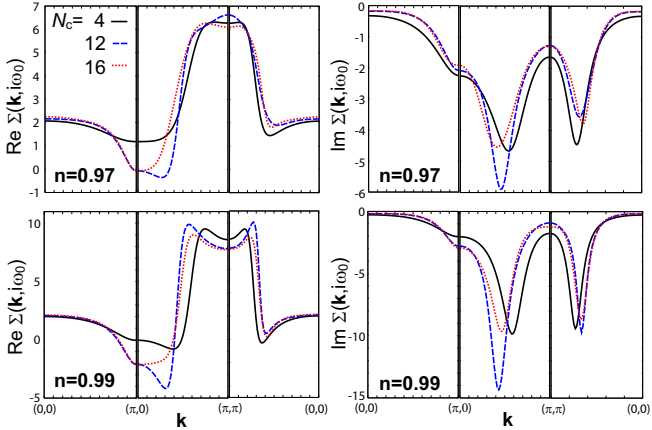}}
\caption{(Color online). Self-energies obtained by the $M$ periodization
         for various clusters at $n=0.97$ and $0.99$.
        }
\label{fig:kdep2}
\end{figure}

We now turn to the quantities periodized through Eqs.~(\ref{eq:periodize}) and (\ref{eq:Gscheme}) (i.e., the physical observables in the full momentum space).
While in the limit of large cluster size all the periodization schemes should give the same result, it depends on the schemes how fast the results converge to the thermodynamic limit with increasing cluster size.
Since the tractable cluster size is rather small at low temperatures, it is important to find out the most efficient periodization scheme in the relevant parameter region.
In this section we compare the self-energies calculated by various periodization schemes for the 2$\times$2, 4$\times$3, and 4$\times$4 clusters, focusing on the slightly hole-doped region.
We present the results for tilted clusters of $N_\text{C}=8$ and 12 in Appendix B.
Since Figs.~\ref{fig:locality}(a)-(c) indicate that the momentum dependence becomes most significant at low energy, we focus on the self-energies at the lowest Matsubara frequency $\w_0=\pi T$.

Figure \ref{fig:kdep} compares the $M$, $\S$, and $G$ periodizations at $n=0.95$.
We first notice that all the results show a common feature that Im$\Sigma$ is small around $(0,0)$ and large around $(\pi,0)$ and $(\pi,\pi)$.
Looking at the convergence against the cluster size, we notice that the $M$-periodized self-energy fluctuates much less than the $\S$- and $G$-periodized ones.
This indicates a smaller size effect in the $M$ periodization, consistently with the results discussed in Secs.~\ref{ssec:ih} and \ref{ssec:locality}.
Roughly speaking, the fluctuation is large around $(\pi,0)$ and $(\pi,\pi)$ while small around $(0,0)$ and $(\frac{\pi}{2},\frac{\pi}{2})$, reflecting the momentum-dependent amplitude of the self-energy.
It is interesting that for the $M$ periodization 4$\times$3 and 4$\times$4 cluster results agree well, in particular, around $(0,0)$ and $(\frac{\pi}{2},\frac{\pi}{2})$, indicating the self-energy is nearly converged there.
It is also worthwhile noting that the 2$\times$2 cluster result with the $M$ periodization already reproduces well the overall structure of the 4$\times$3 and 4$\times$4 cluster results.

The bottom panels in Fig.~\ref{fig:kdep} compare the 4$\times$3 and 4$\times$4 cluster results with the $M$, $\S$, and $G$ periodizations.
We see a nice agreement around $(0,0)$ and $(\frac{\pi}{2},\frac{\pi}{2})$, especially between the $M$ and $G$ schemes.
The agreement between the different periodization schemes corroborates our expectation for the convergence.

Figure \ref{fig:kdep2} presents the $M$-periodized self-energy at $n=0.97$ and 0.99.
The agreement between the 4$\times$3 and 4$\times$4 cluster results is still nice around $(0,0)$ and $(\frac{\pi}{2},\frac{\pi}{2})$ even at these small dopings.
Also the 2$\times$2 cluster result looks well reproducing these larger-cluster results. 

\begin{figure}[t]
\center{
\includegraphics[width=0.48\textwidth]{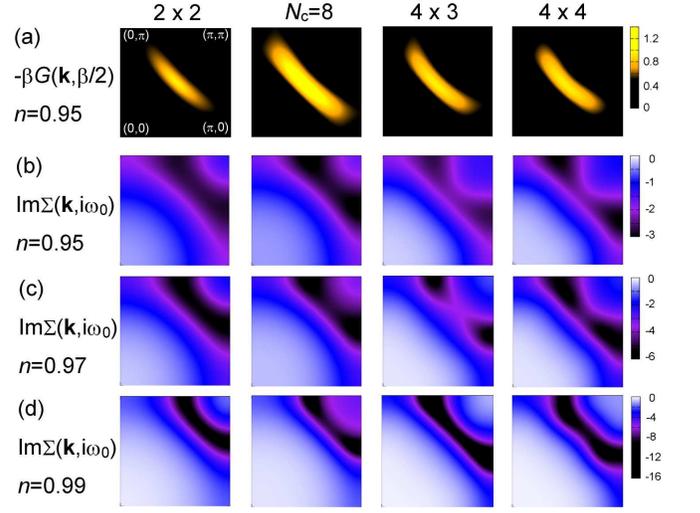}}
\caption{(Color online). (a) Momentum maps of $-\b G(\Vec{k},\frac{\b}{2})$               obtained through the $M$ periodization for the chosen cluster sizes 
          at $n=0.95$.
          (b)-(c) Those of Im$\S(\Vec{k},i\w_0)$ at $n=0.95$, 0.97, and 0.99.
          Here the results with $N_\text{C}=8$ are also presented for comparison.        }
\label{fig:kmap}
\end{figure}

Figure \ref{fig:kmap}(a) is the momentum map of the $M$-periodized Green's functions at $\tau=\frac{\b}{2}$ for the employed cluster sizes at $n=0.95$.
The quantity gives an estimate of the low-energy spectral weight because  
\begin{align}\label{eq:ghb}
-\b G(\Vec{k},\t=\beta/2)=\frac{\b}{2}
\int_{-\infty}^{\infty}\frac{A(\Vec{k},\w)}{\cosh(\b\w/2)}d\w
\end{align}
is approximately a spectral weight averaged over an energy width $\sim T$ around $\w=0$. Hence the momentum map indicates the shape of the Fermi surface.
The results show the Fermi arc structure, found previously in 2$\times$2 CDMFT studies,\cite{st04,cc05,sk06,c09,sm09,sm10}, for all the cluster sizes.
Remarkably the location of the arc does not change significantly with the cluster size.
This is in agreement with the fast convergence of the self-energy around the nodal point seen in Fig.~\ref{fig:kdep}.
Although we have not detected any indication of the Fermi pocket at this temperature ($T=0.06t$), it is still consistent with the pocket structure at $T=0$, as discussed in Refs.~\onlinecite{sm09} and \onlinecite{sm10}.

Figures \ref{fig:kmap}(b)-(d) are the momentum maps of the imaginary part of the $M$-periodized self-energy at $n=0.95$, 0.97, and 0.99.
For all the cluster sizes and dopings Im$\S$ shows a strong intensity around the cut along $(\frac{\pi}{2},\pi)-(\pi,\frac{\pi}{2})$.
The large $|\text{Im}\S|$ indicates the presence of a zero surface of Green's function at $T=0$.
The indicated location of the zero surface for the 2$\times$2 cluster is 
consistent with that previously found by 2$\times$2 CDMFT+ED studies,\cite{sk06,sm09} while it deviates from the one assumed in Ref.~\onlinecite{yr06}.
It is remarkable that the location of the zero surface does not change significantly with the cluster size.
It seems hardly moved with reducing doping from 5\% to 1\% as well.
This is quite unexpected from the viewpoint of the extended Luttinger sum rule \cite{d03} which argues that the sum of the volume enclosed by the Fermi surface and by the zero surface is equal to the electron filling.
If this rule holds, such a volume should be much smaller than the one indicated  in Fig.~\ref{fig:kmap} because the volume should approach to half of the full Brillouin zone as the filling approaches to the half filling.
Figure \ref{fig:kmap} indicates an anomalous metallic phase characterized by the simultaneous presence of both zero and Fermi surfaces. This phase is separated by a quantum phase transition from the Fermi liquid,\cite{sm09} which we find at high doping and which appears, within our numerical precision, to fully respect the Luttinger sum rule.

There have been intensive debates on the applicability of the Luttinger sum rule in strongly correlated region.\cite{r07,f07}
Although the present results strongly suggest that the sum rule is violated in the underdoped region, the cluster-size dependence of $\Sigma$ remaining around $(\pi,0)$ and $(\pi,\pi)$ makes difficult to make definitive statements.
Future studies on larger clusters are highly desired to settle this issue.

The present study indicates that many of the characteristic features of the doped Mott insulators summarized as the momentum differentiation and identified in the pseudogap and Fermi arc formation are well captured even at the temperature scale of 0.06$t$ or above. In this range of temperature, the appropriate $M$ periodization scheme allows the convergence to the thermodynamic limit at a relatively small cluster size.  If one wishes to see the growth of the electron coherence with higher energy resolution close to the Fermi level, one needs to go to lower temperatures together with the corresponding larger cluster size. Increasing the resolution in the momentum space simultaneously at lower temperatures beyond the present study is a challenge left for future studies.

\section{SUMMARY AND CONCLUSION}\label{sec:summary}
We have extended the cellular DMFT to clusters larger than the conventionally used 2$\times$2 one, and systematically studied the cluster-size dependence of various quantities.
While the CDMFT sacrifices the translational symmetry of the original lattice, it can provide through a periodization a fine structure in the momentum space.
This is an interesting information, complementary to DCA results, where
the self-energy is assumed to be flat in each momentum patch.\cite{footnote2}
Our strategy is to find and use an efficient quantity for the periodization, to be able to extract the thermodynamic behavior from a relatively small cluster calculation.

In order to achieve this task, we have explored how homogeneous and local the various cluster quantities are.
Focusing on the parameter region of hole-doped Mott insulators, we have found that the cluster cumulant is the most local and homogeneous quantity, favorable for the periodization.
The comparison of the self-energies obtained by various periodization schemes shows that the fastest convergence against cluster size is obtained by the periodization of the cumulant $M$.
The convergence depends on momentum: While it seems converged already at the 4$\times$4 cluster around $(0,0)$ and $(\frac{\pi}{2},\frac{\pi}{2})$, a distinct size dependence still remains around $(\pi,0)$ and $(\pi,\pi)$.
We have also found that the 2$\times$2 cluster with the $M$ periodization remarkably well reproduces the overall structure of the self-energy obtained with the  4$\times$3 and 4$\times$4 clusters.
The Fermi arc structure and the location of the low-energy zero surface, calculated through the $M$ periodization, seem only weakly dependent on the cluster size, corroborating the picture of the Mott physics obtained by previous CDMFT studies.
This result would imply a violation of the extended Luttinger theorem at small doping.

\section*{ACKNOWLEDGMENT}
We thank A. J. Millis, E. Gull, and N. Lin for valuable comments. 
M. C. thanks T. D. Stanescu for fruitful discussions.
This work was supported by Grant-in-Aid for Scientific Research
 (Grant No. 22340090), from MEXT, Japan. 
A part of this research has been funded by the Strategic Programs for 
Innovative Research (SPIRE), MEXT, and the Computational Materials 
Science Initiative (CMSI), Japan. 
S. S. is supported by the Japan Society for the Promotion of Science 
for Young Scientists, and K. H. by 
the Austrian Science Fund (FWF) through SFB ViCoM F4103-N13.
G. S. acknowledges support from the FWF under ``Lise-Meitner" Grant No. M1136.
The calculations have been performed at Vienna Scientific Cluster, at Information Technology Center, Nagoya University, and at the Supercomputer Center, ISSP, University of Tokyo.

\section*{APPENDIX A: High-frequency expansion and relation between $M$, $\Sigma$, and $G$ periodizations}

We first give a simple argument why the correct implementation of the $G$ periodization scheme is given by Eq.~(\ref{eq:Gscheme}) rather than by Eq.~(\ref{eq:periodize}) with $\hat{Q}^\text{C}=\hat{G}^\text{C}$.
\begin{widetext}

The two Green's functions $\hat{G}(\widetilde{\Vec{k}},z)$ and $\hat{G}^\text{C}(z)$ defined in Eqs.~(\ref{eq:selfcon}) and (\ref{eq:Gc}), respectively, can be expanded for large values of $z=i\omega_n+\mu$:
\begin{equation}\label{Gk1}
\hat{G}(\widetilde{\Vec{k}},z) 
= \left( z\hat{I} -\hat{t}(\widetilde{\Vec{k}}) -\hat{\Sigma}^\text{C}(z)  \right)^{-1} 
= \frac{1}{z}\hat{I} + \frac{1}{z^2}\left( \hat{t}(\widetilde{\Vec{k}}) +\hat{\Sigma}^\text{C}(z)  \right) + \cdots,
\end{equation}
\begin{equation}\label{GC1}
\hat{G}^\text{C}(z) 
= \frac{N_\text{C}}{(2\pi)^2} \int_\text{RBZ} d\widetilde{\Vec{k}} \hat{G}(\widetilde{\Vec{k}},z) 
=\frac{N_\text{C}}{(2\pi)^2}\int_\text{RBZ} d\widetilde{\Vec{k}} \left( z\hat{I} -\hat{t}(\widetilde{\Vec{k}}) -\hat{\Sigma}^\text{C}(z)  \right)^{-1}  
= \frac{1}{z}\hat{I} + \frac{1}{z^2}\left(
\frac{N_\text{C}}{(2\pi)^2}\int_\text{RBZ} d\widetilde{\Vec{k}} \hat{t}(\widetilde{\Vec{k}}) +\hat{\Sigma}^\text{C}(z)  \right) + \cdots.
\end{equation}
Let us note that, for the present analysis of the different periodization schemes, the explicit $1/z$ expansion of $\hat{\Sigma}^\text{C}(z)$ is not necessary.

Written in this form, it is simple to see what happens if one periodizes $\hat{G}(\widetilde{\Vec{k}},z)$ and $\hat{G}^\text{C}(z)$, that is, if one applies Eqs.~(\ref{eq:Gscheme}) and (\ref{eq:periodize}):
\begin{equation}\label{Gk2}
G^\text{L}(\Vec{k},z)|_{\text{from }\hat{G}(\widetilde{\Vec{k}},z)} 
= \frac{1}{N_\text{C}} \sum_{i,j\in \text{C}} \operatorname{e}^{i \Vec{k} \cdot (\mathbf{r}_i-\mathbf{r}_j)} G_{ij}(\widetilde{\Vec{k}},z) 
= \frac{1}{z} + \frac{1}{z^2}
\frac{1}{N_\text{C}} \sum_{i,j\in \text{C}} \operatorname{e}^{i {\Vec{k}} \cdot (\mathbf{r}_i-\mathbf{r}_j)} \left( 
t_{ij}(\widetilde{\Vec{k}}) +\Sigma_{ij}^\text{C}(z)  \right) + \cdots,
\end{equation}
and
\begin{equation}\label{GC2}
G^\text{L}({\Vec{k}},z)|_{\text{from }\hat{G}^C(z)} 
= \frac{1}{N_\text{C}} \sum_{i,j\in \text{C}} \operatorname{e}^{i \Vec{k} \cdot (\mathbf{r}_i-\mathbf{r}_j)} G_{ij}^\text{C}(z) 
= \frac{1}{z} + \frac{1}{z^2}  \frac{1}{N_\text{C}} \sum_{i,j\in \text{C}} \operatorname{e}^{i \Vec{k} \cdot (\mathbf{r}_i-\mathbf{r}_j)} \left(\frac{N_\text{C}}{(2\pi)^2} \int_\text{RBZ} d\widetilde{\Vec{k}}  \ t_{ij}(\widetilde{\Vec{k}}) + \Sigma_{ij}^\text{C}(z)  \right) + \cdots.
\end{equation}

Using the definition given just below Eq.~(\ref{eq:selfcon}), we can re-express the term involving $\hat{t}(\widetilde{\Vec{k}})$ in Eq.~(\ref{Gk2}) as
\begin{equation}\label{Gk3}
 \frac{1}{N_\text{C}} \sum_{i,j\in \text{C}} \operatorname{e}^{i {\Vec{k}} \cdot (\mathbf{r}_i-\mathbf{r}_j)} t_{ij}(\widetilde{\Vec{k}}) = 
\frac{1}{N_\text{C}} \sum_{i,j\in \text{C}} \operatorname{e}^{i {\bf k} \cdot (\mathbf{r}_i-\mathbf{r}_j)}  
 \sum_{\mathbf{K}} e^{-i(\ksmall+\mathbf{K})\cdot(\mathbf{r}_i-\mathbf{r}_j)} \epsilon(\ksmall+\mathbf{K}) = \epsilon(\mathbf{k}),
\end{equation}
and that in Eq.~(\ref{GC2}) as
\begin{equation}\label{GC3}
 \frac{1}{N_\text{C}} \sum_{i,j\in \text{C}} \operatorname{e}^{i \Vec{k} \cdot (\mathbf{r}_i-\mathbf{r}_j)} \frac{N_\text{C}}{(2\pi)^2}
\int_\text{RBZ} d\widetilde{\Vec{k}}\  t_{ij}(\widetilde{\Vec{k}}) =
\frac{1}{N_\text{C}} \sum_{i,j\in \text{C}} \operatorname{e}^{i \Vec{k} \cdot (\mathbf{r}_i-\mathbf{r}_j)} t_{ij}^\text{C} \neq \epsilon(\mathbf{k}),
\end{equation}
where $\hat{t}^\text{C}$ denotes the hopping matrix restricted to the cluster.
The left-hand side of Eq.~(\ref{GC3}) is different from the full lattice dispersion $\epsilon(\mathbf{k})$ that one would like to be reproduced in the lattice Green's function at the second order in the $1/z$ expansion.
Such a simple argument thus shows that the periodization of the cluster Green's function matrix [i.e., Eq.~(\ref{GC2})] is not an appropriate periodization scheme. 

\end{widetext}

Conversely one can easily show that at the second order in the $1/z$ expansion the $M$- and $\Sigma$-periodized Green's functions yield the same result as Eq.~(\ref{Gk2}), and in particular they reproduce the right band dispersion.
The $G$-, $M$-, and $\Sigma$-periodization schemes differ instead already at the third order in $1/z$.
In the $G$ periodization the third order term in the $1/z$ expansion reads
\begin{equation}
 \frac{1}{z^3} \frac{1}{N_\text{C}} \sum_{i,j\in \text{C}}
      \operatorname{e}^{i {\Vec{k}} \cdot (\mathbf{r}_i-\mathbf{r}_j)} 
\left[ \left(\hat{t}(\widetilde{\Vec{k}}) + \hat{\Sigma}^\text{C} \right)^2
\right]_{ij},
\end{equation}
while in the $M$ periodization it is
\begin{equation}
\frac{1}{z^3} \left[ \e^2(\Vec{k}) 
   + \frac{1}{N_\text{C}} \sum_{i,j\in \text{C}} 
     \operatorname{e}^{i {\Vec{k}} \cdot (\mathbf{r}_i-\mathbf{r}_j)}
     \left\{ 2\e(\Vec{k}) \hat{\Sigma}_{ij}^\text{C}
            +\left[\left(\hat{\Sigma}^\text{C}\right)^2\right]_{ij} \right\} \right],
\end{equation}
and in the $\Sigma$ periodization it is
\begin{equation}
\frac{1}{z^3} \left[ \e(\Vec{k}) 
   + \frac{1}{N_\text{C}} \sum_{i,j\in \text{C}} 
     \operatorname{e}^{i {\Vec{k}} \cdot (\mathbf{r}_i-\mathbf{r}_j)}
     \hat{\Sigma}_{ij}^\text{C} \right]^2.
\end{equation}
We notice that the $M$ and $G$ periodizations share the same term proportional to $(\hat{\Sigma}^\text{C})^2$ (while $M$ and $\Sigma$ periodizations share the term $\e_{\Vec{k}} \S_{\Vec{k}}$).
Therefore, in a strongly correlated regime where the self-energy is large compared to the bare dispersion, we may expect that the $M$ and $G$ schemes produce a similar $G^\text{L}$, while the $\S$-periodized Green's function should deviate from it. This is indeed observed in our CDMFT solution for doped Mott insulators, where the self-energy is large (see bottom panel in Fig.~\ref{fig:kdep}).

These considerations can be also clarified by expanding the $G$-periodized Green's function [Eq.~(\ref{eq:Gscheme})] with respect to the cumulant
(if the self-energy is large compared to the bare hopping terms, $\hat{M}^\text{C}\hat{t}$ is small):
\begin{eqnarray}
&&\hat{G}(\widetilde{\mathbf{k}}, i\omega_n) 
=  \left[ \hat{I}- \hat{M}^\text{C}( i\omega_n) \hat{t} (\widetilde{\mathbf{k}}) \right]^{-1} \hat{M}^\text{C}(i\omega_n) \nonumber\\
&=&
\hat{M}^\text{C}(i\omega_n) + \hat{M}^\text{C}(i\omega_n) \hat{t}(\widetilde{\mathbf{k}}) \hat{M}^\text{C}(i\omega_n) + \cdots. 
\label{eq:expansion}
\end{eqnarray}
Hence, $G^{\text{L}}(\Vec{k},i\w_n)|_{\text{from }\hat{G}(\widetilde{\Vec{k}},i\w_n)}\equiv M^\text{L}(\Vec{k},i\w_n)$ at the first order.

On the other hand, in the $M$ periodization, the lattice Green's function is given by Eq.~(\ref{eq:G_M}), so that 
\begin{eqnarray}
&&G^{\text{L}}(\Vec{k},i\w_n)|_{\text{from }\hat{M}^\text{C}(i\w_n)} = \frac{M^\text{L}(\Vec{k},i\omega_n)}{1- M^\text{L}(\Vec{k},i\omega_n) \e(\Vec{k})} \nonumber\\
&=& M^\text{L}(\Vec{k},i\w_n) 
+M^\text{L}(\Vec{k},i\w_n) \e(\Vec{k}) M^\text{L}(\Vec{k},i\w_n)+ \cdots. 
\label{eq:expansion_GMk}
\end{eqnarray}
Thus $G^{\text{L}}(\Vec{k},i\w_n)|_{\text{from }\hat{G}(\widetilde{\Vec{k}},i\w_n)} \equiv G^{\text{L}}(\Vec{k},i\w_n)|_{\text{from }\hat{M}^\text{C}(i\w_n)}$ at the first order in $\hat{M}^\text{C} \hat{t}$, and we expect that the $M$ and $G$ periodizations produce similar results close to the insulating state, even at energy scales lower than the range of validity of the $1/z$ expansion.
We stress that the lattice Green's function obtained with the $\S$ periodization gives a result {\it a priori} very different. 
This is because the periodization of $\hat{\S}^\text{C}(i\omega_n)=i\omega_n+\mu-\hat{M}^\text{C}(i\omega_n)^{-1}$ corresponds to the periodization of $\hat{M}^\text{C}(i\omega_n)^{-1}$, which gives a lattice cumulant very different from the one obtained with the direct periodization of the cluster cumulant $\hat{M}^\text{C}(i\omega_n)$.

\section*{APPENDIX B: EFFECT OF CLUSTER GEOMETRY}

\begin{figure}[t]
\center{
\includegraphics[width=0.48\textwidth]{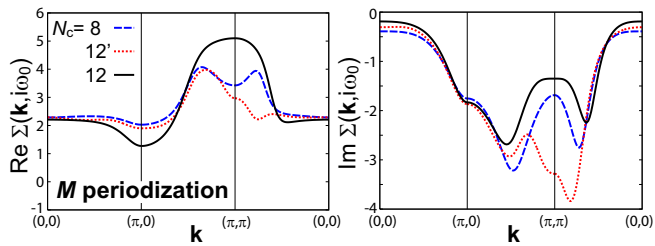}
}
\caption{(Color online). Self-energies obtained by $M$ periodization
         at $n=0.95$ for tilted clusters of $N_\text{C}=8$ and 12.
         For comparison the result for the 4$\times$3 cluster is also plotted.        }
\label{fig:kdep3}
\end{figure}

In general we still have several options in the cluster shape even when the cluster size is fixed.
The cluster shape significantly affects the inhomogeneity of cluster quantities since each cluster site is differently connected to the other cluster sites and/or bath sites.
For example, ``surface" sites on the boundary of the cluster behave differently from the inner sites.
Since nonlocal correlations will be more accurately incorporated in the inner-site self-energy \cite{bk05} than in the surface-site one, a strategy to choose the cluster shape is to minimize the effect of the surface sites.

This was already demonstrated in Figs.~\ref{fig:ih_local} and \ref{fig:ih_nn}. 
When comparing the results at the corner site 1 and at the other surface site 2, the former deviates more from those at the inner site 6 than the latter. 
This reflects the fact that site 1 has only two nearest neighbors in the cluster, while site 2 has three neighbors.

Here we compare the two different 12-site clusters shown in Fig.~\ref{fig:cluster}(a).
An important difference is that the $N_\text{C}=12'$ cluster has two tip sites coupling to only one nearest neighbor while the 4$\times$3 cluster has no such sites.
The cluster cumulants associated with these tip sites indeed show a peculiar behavior largely deviated from the inner ones (not shown).

The difference is reflected in the $M$-periodized self-energy plotted in Fig.~\ref{fig:kdep3}, where we see a large deviation around $(\pi,0)$ and $(\pi,\pi)$ while a fairly nice agreement around $(0,0)$ and $(\frac{\pi}{2},\frac{\pi}{2})$.
We see that the $N_\text{C}=12'$ result is rather closer to the $N_\text{C}=8$ one, which also has two tip sites.

The result indicates that we should avoid clusters containing a tip site, in order to obtain a fast convergence to the thermodynamic limit.
A simple prescription for this is to take a square cluster whose edges are parallel to the lattice vectors, as we have done so far.



\begin{references}
\bibitem{footnote}
For $N_\text{C}=8$ and 12, we average over $x$ and $y$ directions in Eqs.~(\ref{eq:periodize}) and (\ref{eq:Gscheme}).
\end{references}
\end{document}